**Analyzing Social Interaction Networks from Twitter for Planned Special Events**

Arif Mohaimin Sadri, Ph.D.
Lyles School of Civil Engineering
Purdue University, West Lafayette, IN 47907, USA
Email: sadri.buet@gmail.com
(Corresponding Author)

Samiul Hasan, Ph.D.
Assistant Professor
Department of Civil, Environmental, and Construction Engineering
University of Central Florida
12800 Pegasus Drive, Orlando, FL 32816
Phone: 407-823-2841
Email: samiul.hasan@ucf.edu

Satish V. Ukkusuri, Ph.D.
Professor
Lyles School of Civil Engineering
Purdue University, West Lafayette, IN 47907, USA
Email: sukkusur@purdue.edu

Juan Esteban Suarez Lopez
School of Civil Engineering
National University of Colombia
Medellin, Colombia
Email: jesuarezl@unal.edu.co

**ABSTRACT**

The complex topology of real networks allows its actors to change their functional behavior. Network models provide better understanding of the evolutionary mechanisms being accountable for the growth of such networks by capturing the dynamics in the ways network agents interact and change their behavior. Considerable amount of research efforts is required for developing novel network modeling techniques to understand the structural properties such networks, reproducing similar properties based on empirical evidence, and designing such networks efficiently. In this study, we first demonstrate how to construct social interaction networks using social media data and then present the key findings obtained from the network analytics. We analyze the characteristics and growth of online social interaction networks, examine the network properties and derive important insights based on the theories of network science literature. We also discuss the application of such networks as a useful tool to effectively disseminate targeted information during planned special events. We observed that the degree-distributions of such networks follow *power*-law that is indicative of the existence of fewer nodes in the network with higher levels of interactions, and many other nodes with less interactions. While the network elements and average user degree grow linearly each day, densities of such networks tend to become zero. Largest connected components exhibit higher connectivity (density) when compared with the whole graph. Network radius and diameter become stable over time evidencing the small-world property. We also observe increased transitivity and higher stability of the power-law exponents as the networks grow. Since the data is specific to the *Purdue University* community, we also observe two very big events, namely *Purdue Day of Giving* and *Senator Bernie Sanders'* visit to *Purdue University* as part of *Indiana Primary Election 2016*.

**BACKGROUND AND MOTIVATION**
Planned Special Events (PSE) include sporting events, concerts, conventions and similar large events at specific venues such as stadiums and convention centers among others. PSEs possess many operational needs that can be anticipated and managed in advance because of specific locations and times of occurrence [1]. Organizing PSEs have several challenges including parking management, crowd management, pedestrian facility design, and special facility for senior citizens and handicapped individuals, providing transit facility for captive riders among others. In addition, police enforcements often need to close several streets for security reasons, manage crowds who walk together to the location and guide motorists to specific routes who are unfamiliar with the area. Individuals attending these events travel by various travel modes, i.e. walk, private car and public transit. Since the traffic patterns of PSEs vary significantly as compared to any given weekday traffic patterns, accidents, or any other incidents, it is of great inconvenience for traffic managers, drivers or freight movers to deal with PSEs [1]. Thus, PSEs are a major concern for traffic planners and local transportation agencies because of increased traffic demand and restricted roadway capacity causing disruptions to the regular traffic conditions [1, 2]. However, this disruption and the associated operational needs can be anticipated and managed in advance [3]. Participation from key stakeholders, development and implementation of effective traffic management plan, and the flexibility to change plans to manage real-time traffic are among the key strategies to efficiently handle PSEs [4].

*Network Science* is an emerging research field having multifaceted outlook in the study of large-scale real networked systems which considers both the network topology and the behavior of network actors. Complex networks with dynamic and irregular structure along with their statistical properties are the primary focus of such research efforts i.e. the combined knowledge of network structure and behavior which is significantly distinct from the straight-forward analysis of single small graphs [5, 6]. The prevalence of networked systems has resulted in a number of studies with applications in various domains over the last decade. Such domains include social organizations, internet, systems biology, supply chains and logistics, information and communication systems, financial markets, infrastructure systems among others [6-8]. A number of novel structural properties and concepts have been observed and some unifying principles along with relevant statistical distributions have been derived to characterize interdependence between the structure and function of complex real networks. For example, the existence of *small-world property* in many real networks suggest that most vertices in the network can be reached from every other node by using relatively short paths despite large size of these networks [9-11]. On the other hand, the *scale-free property* suggests the existence of large hubs i.e. a few nodes which are highly connected to many other nodes in the network and such hubs result in a network degree distributions (*power-law*) with highly right-skewed long tail referring to nodes with a much higher degree than most other nodes [12]. Other properties include transitivity, network resilience, mixing patterns, network homophily or similarity, degree correlations, preferential attachment, community structure, network navigation, size of largest components among other which provide valuable insights [6].

Information dissemination is the systematic way of distributing information and spreading awareness to every individual and systematic planning, collection, organization, and delivery technique are needed before circulating relevant information to any target audience by using various media and communication means. Information dissemination thus constitutes an important and critical factor for the success of organizing PSEs. Despite many technical challenges to manage PSEs, the empirical literature does not provide any specific guidelines to local traffic managers and emergency response personnel to disseminate travel specific information as part of traffic management plans for PSEs. Social media platforms (Twitter, Facebook and others) can be considered as appropriate means of disseminating information dynamically. Studies have found that an individual's real world actions can be inferred based on the connections and activities in social media [13]. Twitter shows both the characteristics of a social network and an informational network [14] and users can share short messages up to 140 characters along with the ability to follow other. While the information network properties of Twitter instigate information contagion globally, the social network properties allows access to geographically and personally relevant information and [15]. Because of specific features, Twitter can be particularly useful for effective information dissemination during PSEs.

The overwhelming usage and activity of Twitter users can be expressed by a 2013 statistic that suggests over 143K tweets per second being generated on Twitter [16]. User activity in social media has shown its prevalence in recent years and the world is more connected now than ever before. The ease in information sharing and the ability to instigate others have primarily contributed to such connectivity all over the world. However, the purpose and context of the online activity on social media platforms, such as Facebook and Twitter, may vary from user to user. For example, users' check-in activity can be referred as distinct from what users' post or share to disseminate any specific information. One specific feature of such information sharing activity is the ability of users' in mentioning (direct mentions, retweets and replies) others that they follow.

In this study, we used Twitter REST API for the keyword 'purdue' and obtained 56,159 tweets over a month. Each tweet consisted of several words and user mentions which co-appeared with the keyword 'purdue' and resembled higher likelihood of a Twitter subscriber belonging to the Purdue University community. Next, we construct the 'purdue' specific online social interaction networks by considering the user mentions. Finally, we analyze such network characteristics, examine the properties and network growth, and derive relevant insights based on the theories of network science literature.

**RELATED WORK**
The coupling between network structure and behavior of network agents has initiated a number of studies related to the dynamical behavior of network actors communicating through complex network topologies. This interdependence has significant outcomes when the robustness and resilience of a real network is considered and the way networks respond to targeted failure due to external disturbances [17]. Fundamental research questions, as explored in the empirical literature, can be listed as (i) how to model emergence of new innovations or ideas based on agent interactions, (iii) who are the key influential players in the network and how to find them, (iii) how to maximize network influence based on certain mechanism, (iv) when and how networks execute contagious behavior such as disease transmission or information propagation, (v) which agents are more likely to connect and interact with agents of the same profile? A few examples of such studies may include epidemic models of disease transmission [18, 19], email networks and computer virus transmission [20, 21], collapse in power grid networks [22, 23], disruptions of trade markets [24], information propagation in social networks [25], and many others.

The successful spreading of awareness to every individual in a community solely depends on an effectual information dissemination process [26-28]. The prevalence of user activities in social media (Twitter and Facebook for example) has shown its prevalence over the last decade, allowing people to be more connected than ever before worldwide. This has been possible because of the flexibility in information sharing and the way people instigate their online neighbors. Unlike the usual communication media, social media can be used as useful means to disseminate information during any specific event such as PSEs. The benefits to convincingly collect, analyze and use such large-scale and rich information from online information sources have also been addressed [29]. Many empirical studies have explored social media data for emergency response and disaster management [30-35], crisis informatics [36-43] and many others. In fact, one could efficiently analyze and predict real world human actions based on user activity and connectivity on social media platforms [13, 15]. However, the purpose and context of the user activity over social media platforms may vary from one user to another user.

Transportation researchers have started exploiting these large-scale data sources more extensively in recent years. Such examples include problems related to travel survey techniques [44, 45], activity-pattern modeling [46-48], origin-destination demand estimation [49-53] and transit planning [54]. However, a few studies explored people's ego-centric offline social networks and the effects of ego-centric social ties on joint activity participation [55, 56]. To the best of the authors' knowledge, the empirical literature does not provide any specific evidence on how to extract networks of user interactions from social media and the applicability of such networks to solve transportation related problems, which is the primary scope of this study.

**NETWORK DATA DESCRIPTION**

We used twitter REST API to collect tweets for using a specific keyword 'purdue' which is frequently used in Twitter within the Purdue community. The dataset selection was due to its relevance in the study of information flow patterns of specific topics related to Purdue University. It is equally important to understand how special events affect this behavior over time. By running the twitter REST API between April 16, 2016 and May 16, 2016 for four consecutive weeks, we were able to obtain 56,159 tweets for the query 'purdue' after initial data cleaning for non-English tweets and common stop words. 19,532 of these tweets did not include any user mentions, however, the rest of the tweets included at least one user mention in each tweet. The frequency distribution of the top 50 words in these tweets, co-appeared with 'purdue', suggests that these top 50 most frequent words contribute up to ~25% of all the words that appeared in the collected data. A word cloud of the most frequent words in the dataset is presented in Figure 1(a). Different combinations of these words constitute specific topics based on which users influence one another by using the user mention feature on Twitter. These frequently appeared words also suggest the emergence of event specific topics such as *Purdue Day of Giving, Senator Bernie Sanders's* visit during *Indiana Primary* among others. Celebrity players of Purdue such as *Anthony Brown* (football), *Spike Albrecht* (basketball) and others also contributed to many topics. The differences in the amount of user mentions in the tweets over days are plotted in Figure 1(b). It can be clearly seen that the number of tweets having user mentions is almost twice as the number of tweets without mentions. These tweets primarily contribute to the formation of networks of direct social influence. In Table 1, we present the amount of user mentions as was observed in the tweets.

In Twitter, users can post tweets up to 140 characters and each data point can be stored as a tuple *Tweet* once collected with the following information:

*Tweet (tweet_id) = {tweet, tweet_created_at, user_id, user_screen_name, user_location, user_name, user_followers_count, user_friends_count, user_statuses_count, user_favourites_count,, user_listed_count, user_mention, tweet_retweeted, tweet_lat, tweet_lon}*

For this study, we are interested in using a sub-tuple *tweet* to infer the links of direct influence that finally evolves into a highly connected network of a given context:

*tweet (tweet_id) = { user_id, tweet, user_mention}*

Let us consider the following three tuples from the tweets generated on Twitter on 04/28/2016 (02:45:40 +0000), 05/02/2016 (14:45:33 +0000) and 05/03/2016 (13:50:21 +0000), respectively.

*tweet(725516302819938305) = {709920419529281537, **'at purdue university, we are in this campaign to win and become the democratic nominee. - bernie sanders htt…'**, [null]}*

*tweet(727147016233558016) = {3239853627, **'rt @saracohennyc at purdue university, we are in this campaign to win and become the democratic nominee. - bernie sanders htt…'**, [709920419529281537]}*

*tweet(727495513277382656) = {325069363, **'rt @bernielovesall: rt @saracohennyc at purdue university, we are in this campaign to win and become the democratic nominee. - bernie sanders htt…'**, [3239853627, 709920419529281537]}*

Based on the above tweets, there is a directed link from user *709920419529281537* to user *3239853627* and from user *709920419529281537* to *325069363*. Please refer to Figure 2 (a) for the details of this network construction. The preliminary analysis of these data waves suggest the existence of 34,363 unique users and 38,442 unique undirected links (39,709 links if direction is considered) of direct influence. The network elements of the graph (constructed based on the data) are presented in Table 1.

Different network visualizations are presented from Figure 2(c)-2(f) to depict the network configurations and how the network structure appears after four weeks. In Figure 2(c), we present the undirected graph having 34,363 users from Twitter and 38,442 links. The network isolates without any connectivity are also shown in the periphery. This graph includes 8,348 connected components and the largest connected component is presented in Figure 2(e)-2(f). While Figure 2(f) better represents the hierarchical structure of the network with the most central node in the center, Figure 2(e) presents weighted edges based on the number of appearance of these links. This weighted graphs help to explain the existence of links having higher strength which also serves as an evidence of higher influence. In Figure 2(d), we present the largest hub i.e. the most central node having the largest degree. It is highly intuitive that the network will be under huge disruption if such node disappears or remain active in cases.

**IMPLICATIONS FROM NETWORK ANALYSES**

The user mentions, observed in the data for four consecutive weeks, construct a social interaction network that includes 34,363 nodes and 39,709 links for the directed graph and 34,363 nodes and 38,442 links for the undirected graph (Table 1, Figure 2c). 6,096 of these nodes appeared as network isolates (nodes without connectivity) in the periphery along with 8,348 connected components (Figure 2c). 33,020 connections (links) among 21,045 users (nodes) were observed in the largest connected component of this network (Figure 2e-2f). The radius and diameter of the largest connected component were observed as 9 and 17, respectively. These are relevant to the *small world property* of complex real networks that refers to the existence of relatively short paths between any pair of nodes in most networks despite their large size. The existence of this property has been observed in many real networks as studied in the empirical literature [9-11]. This property has significant implications in the modeling of dynamic processes occurring on real networks. For example, when effective information dissemination is considered, contagion will be faster through the network because of short average path lengths [6]. Three important measures to explain this property are eccentricity, radius and diameter. While the eccentricity of a node in a graph is the maximum distance (number of steps or hops) from that node to all other nodes; radius and diameter are the minimum and maximum eccentricity observed among all nodes, respectively.

*Network Density*, frequently used in the sociological literature [57], equals to 0 for a graph without any link between nodes and 1 for a completely connected graph. The density of real graphs refer to the proportion of links that exist in the graph and the maximum number of possible links in the graph. For *n* users, the number of maximum links are *n(n-1)* for a directed graph and *n(n-1)/2* for a undirected graph. The densities that we observe in the social interaction network of 21,045 users are 0.00003, 0.00007 and 0.00015 for the directed, undirected and the largest connected component, respectively. This implies higher connectivity in the largest connected component, more than twice as much as in the whole network. The node *Degree* is the number of edges adjacent to that node, *In-degree* is the number of edges pointing in to the node and *Out-degree* is the number of edges pointing out of the node. The degree of a node ($k$) is the number of direct edges to other nodes in a graph from that node and the degree distribution $P(k)$ in real networks, (probability that a randomly chosen node has degree $k$), is significantly different from the Poisson distribution, typically assumed in the modeling of random graphs. Real networks, in fact, exhibit a power law (or *scale-free*) degree distribution characterized by higher densities of triangles such as cliques in a social network [12]. Such networks also demonstrate significant correlations in terms of node degrees or node-level attributes. Existence of hubs i.e. a few nodes that are highly connected to other nodes, in the network can also be validated by the *scale-free* phenomenon. The largest hub (or ego), as was observed in our dataset, is visualized in Figure 2(d). The presence of large hubs results in a degree distribution with long tail (highly right-skewed), indicating the presence of nodes with a much higher degree than most other nodes. For an undirected network, the degree distribution $P_{degree}(k)$ can be written as follows:

$$P_{degree}(k) \propto k^{-\gamma} \quad \ldots\ldots\ldots\ldots\ldots\ldots\ldots\ldots\ldots\ldots (1)$$

where, $\gamma$ is some exponent and $P_{degree}(k)$ decays slowly as the degree $k$ increases, increasing the probability of obtaining a node with a very high degree. Networks with power-law distributions are called scale-free networks that holds the same functional form (power laws) at all scales. The power law $P_{degree}(k)$ remains unchanged (other than a multiplicative factor) when rescaling the independent variable $k$ by satisfying:

$$P_{degree}(xk) = x^{-\gamma} P_{degree}(k) \quad \text{...........................} (2)$$

The presence of hubs that are orders of magnitude larger in degree than most other nodes is a characteristic of power law networks. The average degree of all the 34,363 users in the social interaction network is 1.156 (Table 1) and the overall degree distributions are plotted in Figure 3(a). Figure 3(b) presents the degree distributions that were observed each day starting from the data period of data collection. By using Alstott et al.'s python code, we obtained the best fitting to the degree distributions [58] and the empirical data of this study fits close to being a power-law or truncated power-law distributions. The code also returns a value of $x_{min}$ which refers to the minimal value of $x$ at which the power law begins to become valid. For power-law, we obtain $\gamma = 2.294 \pm 0.046$; $x_{min} = 11$ and for truncated power-law $\gamma = 2.278$; $x_{min} = 11$. Here, $\gamma$ is the slope of the distribution. When $\gamma$ is high, the number of nodes with high degree is smaller than the number of nodes with low degree. A low value of $\gamma$ may refer to a more equal distribution, whereas higher values of $\gamma$ may denote more and more unfair degree distributions. It is important to note here that the best fit power law may only cover a portion of the distribution's tail [58]. From Figure 3(d), it appears that the data also fits close to being a log-normal distribution. However, difficulties in distinguishing the power law from the lognormal are common and well-described, and similar issues apply to the stretched exponential and other heavy-tailed distributions [59, 60]. Our analysis on the distributions fitting are based on pairwise comparison between power-law, truncated power-law, log-normal, and exponential distributions. See Figure 3(c)-3(d) for details.

Another network property is *Transitivity* that implies the higher likelihood of any two given nodes in a network to be connected, given each of these two nodes are connected to some other node. This property refers to the fact that the friend of one's friend is likely also to be the friend of that person in case of social networks and this is a distinctive deviation from the properties of random graphs. In fact, this is indicative of heightened number of triangles (sets of three nodes each of which is connected to each of the others) that exist in real networks [6] (Newman, 2003a). The existence of triangles can be quantified by *Clustering Coefficient. C:*

$$C = \frac{3* \text{ Number of triangles in the network}}{\text{Number of connected triples of nodes}} \quad \text{................} (3)$$

A *connected triple* refers to a single node with links running to an unordered pair of others. In case of an unweighted graph, the clustering coefficient ($cc_i$) of a node $i$ refers to the fraction of possible triangles that exist through that node:

$$cc_i = \frac{2 T_i}{deg_i*[deg_i-1]} \quad \text{.........................................} (4)$$

Here, $T_i$ is the number of triangles that exist through node $i$ and $deg_i$ is the degree of node $i$. The average clustering coefficient in the undirected social interaction network was observed to be 0.149 (Table 1)

Turning to the results obtained from the network growth analysis, we present these results in Figure 4 and Figure 5. The unit of time for the analysis of network growth was set to be 24-hours. We observe that the growth of network elements [nodes and links in Figure 4(a), isolates and connected components in Figure 4(b)] is almost linear over days except for the date 04/26/2016 for which we could not obtain any data. One key insight here is that the growth rate is higher for days that followed special events such as *Purdue Day of Giving* and *Senator Bernie Sanders'* visit to *Purdue University* during *Indiana Primary 2016*. While the network elements grow linearly, the densities tend to go down to zero because of less overall connectivity in the network (Figure 4c). However, the density of the largest connected components remain slight higher over time. In addition, as the social interaction network keeps growing based on difference in user interaction for various topics, the diameter and radius keeps fluctuating initially, however becomes constant later. This is indicative of network stability when the reachability from one node to another node is considered (Figure 5a-5b). The average degree of the nodes shows similar pattern to that of the growth of network elements initially, however becomes flat later (Figure 5c). The network transitivity, based on average clustering coefficient, suggests that the network becomes more transitive over time, however slight fluctuation still remains (Figure 5d). The power-law exponents each day are presented in Figure 5e. After reducing slightly in the initial days, they turn to becoming flat and take a value close to 2.3. This implies that the power-law property holds when social interaction network is observed over a long period of time.

Finally, we present the existence of repetitions in terms of how elements of such networks appear in the network data (Figure 6). This is of great significance within the context of finding highly active nodes (users) in the social interaction networks along with the strength of relationships between node pairs. The relevance of considering the dynamic strength of social ties in information spreading has been duly addressed [61, 62]. The weighted graph, based on the number of times a link has appeared, is presented in Figure 2(e). In order to assess the commonalities of network elements (nodes and links) over time, we compute the fraction of nodes and links every day that appeared at least once in any of the previous days. From Figure 6(a), it can be seen that 65.2% of the total users (or nodes) on May 16, 2016 appeared in the data at least once. Similarly, we observe that 28.3% of the total links of interaction (undirected graph) on May 16, 2016 appeared, in any of the previous days, at least once (Figure 6b).

**CONCLUSIONS AND KEY FINDINGS**
Real networks having complex topologies demonstrate a unique interdependence between the structure and functional behavior. In this study, we demonstrate such interdependence by exploiting online social interaction networks based on network data obtained from Twitter. The social interaction network was formed by following the user mentions appeared in the tweets during four consecutive weeks which are specific to a university community. The network characteristics and properties have been analyzed and the network growth has been monitored over time. Key insights obtained from the network analyses are listed below:

- The network degree distributions exhibit a *power-law* which is indicative of the *scale-free* property of most real networks. This property holds for any given day as evident from the empirical data. This is indicative of the existence of fewer nodes in the network with higher levels of interactions, and many other nodes with less interactions.
- Network visualization is indicative of some nodes (users) being highly active, some links (relations) having higher strength, existence of network isolates, connected components,

- and hubs i.e. nodes having reachability to many other nodes. This is also evident when the appearance of the network elements each day is compared to all previous days.
- Network elements and average user degree grow linearly each day, however, network densities tend to become zero. Largest connected components exhibit higher connectivity (density) when compared with the whole graph.
- Network radius and diameter become stable over time which suggests less variations when the reachability from one node to another node is considered. These variables are related to the *small-world* property.
- Increased transitivity in the growth of such networks is observed following the pattern of mean clustering coefficient. Initial fluctuations of the power-law exponents reduce as the network grows.

The properties of social interaction networks, as observed in this study, have fundamental implications towards effective information dissemination. For example, power-law degree distributions is related to the resiliency of a communication network. The level of resilience, when a random nodes in the network are removed, depends solely on the way the network is formed i.e. network topology. In case of networks having many low-degree nodes would have less disruption and higher resilience since this nodes lie on few paths between others. However, removal of hubs (high degree nodes) would cause major disruption and network agents would fail to communicate since the regular length of path will increase as a result of many disconnected pairs of nodes. For any *Planned Special Event (PSE)*, the assembling of vehicles and pedestrians in a short amount of time cause transportation and transit authorities to often encounter significant challenges in controlling the induced traffic coming from different origins before the event and departing from the event location after the event. There is hardly any specific method in the empirical literature that would allow local emergency managers or agencies to properly disseminate targeted information to any specific audience as part of traffic management procedures for PSEs. A better knowledge of social interaction network growth and properties would be worthwhile to be considered for such events.

**Acknowledgements**
The authors are grateful to National Science Foundation for the grant CMMI-1131503 and CMMI-1520338 to support the research presented in this paper. However, the authors are solely responsible for the findings presented in this study.

**Author Contributions Statement**
All the authors have contributed to the design of the study, conduct of the research, and writing the manuscript.

**Additional Information**

**Competing financial interests:** Authors declare no competing financial interests.

**REFERENCES**
1. Skolnik J, Chami R, Walker M. Planned Special Events–Economic Role and Congestion Effects. 2008.
2. Carson JL, Bylsma RG. Transportation planning and management for special events2003.
3. Latoski SP, Dunn WM, Wagenblast B, Randall J, Walker MD. Managing travel for planned special events: final report. 2003.
4. Dunn Jr W. Traffic Management of Special Events: The 1986 US Open Golf Tournament. Transportation Research Circular. 1989;(344).
5. Albert R, Barabási A-L. Statistical mechanics of complex networks. Reviews of modern physics. 2002;74(1):47.
6. Newman ME. The structure and function of complex networks. SIAM review. 2003;45(2):167-256.
7. Boccaletti S, Latora V, Moreno Y, Chavez M, Hwang D-U. Complex networks: Structure and dynamics. Physics reports. 2006;424(4):175-308.
8. Ukkusuri SV, Mesa-Arango R, Narayanan B, Sadri AM, Qian X. Evolution of the Commonwealth Trade Network. International Trade Working Paper 2016/07, Commonwealth Secretariat, London. 2016.
9. Milgram S. The small world problem. Psychology today. 1967;2(1):60-7.
10. Travers J, Milgram S. An experimental study of the small world problem. Sociometry. 1969:425-43.
11. Watts DJ, Strogatz SH. Collective dynamics of 'small-world' networks. nature. 1998;393(6684):440-2.
12. Barabási A-L, Albert R. Emergence of scaling in random networks. science. 1999;286(5439):509-12.
13. Korolov R, Peabody J, Lavoie A, Das S, Magdon-Ismail M, Wallace W, editors. Actions are louder than words in social media. Proceedings of the 2015 IEEE/ACM International Conference on Advances in Social Networks Analysis and Mining 2015; 2015: ACM.
14. Myers SA, Sharma A, Gupta P, Lin J, editors. Information network or social network?: the structure of the twitter follow graph. Proceedings of the 23rd International Conference on World Wide Web; 2014: ACM.
15. Kryvasheyeu Y, Chen H, Obradovich N, Moro E, Van Hentenryck P, Fowler J, et al. Rapid assessment of disaster damage using social media activity. Science advances. 2016;2(3):e1500779.
16. Krikorian R. New tweets per second record, and how. Twitter Engineering Blog. 2013;16.
17. Albert R, Jeong H, Barabási A-L. Error and attack tolerance of complex networks. nature. 2000;406(6794):378-82.
18. Anderson RM, May RM, Anderson B. Infectious diseases of humans: dynamics and control: Wiley Online Library; 1992.
19. Murray JD. Mathematical biology I: an introduction, Vol. 17 of interdisciplinary applied mathematics. Springer, New York, NY, USA; 2002.
20. Balthrop J, Forrest S, Newman ME, Williamson MM. Technological networks and the spread of computer viruses. Science. 2004;304(5670):527-9.
21. Newman ME, Forrest S, Balthrop J. Email networks and the spread of computer viruses. Physical Review E. 2002;66(3):035101.

Sadri, Hasan, Ukkusuri, Suarez Lopez                                                                                                          12

22. Kinney R, Crucitti P, Albert R, Latora V. Modeling cascading failures in the North American power grid. The European Physical Journal B-Condensed Matter and Complex Systems. 2005;46(1):101-7.
23. Sachtjen M, Carreras B, Lynch V. Disturbances in a power transmission system. Physical Review E. 2000;61(5):4877.
24. Sornette D. Why stock markets crash: critical events in complex financial systems: Princeton University Press; 2009.
25. Coleman JS, Katz E, Menzel H. Medical innovation: A diffusion study: Bobbs-Merrill Co; 1966.
26. Cutter SL, Finch C. Temporal and spatial changes in social vulnerability to natural hazards. Proceedings of the National Academy of Sciences. 2008;105(7):2301-6.
27. Helbing D. Globally networked risks and how to respond. Nature. 2013;497(7447):51-9.
28. Vespignani A. Predicting the behavior of techno-social systems. Science. 2009;325(5939):425-8.
29. Lazer D, Pentland AS, Adamic L, Aral S, Barabasi AL, Brewer D, et al. Life in the network: the coming age of computational social science. Science (New York, NY). 2009;323(5915):721.
30. Bagrow JP, Wang D, Barabasi A-L. Collective response of human populations to large-scale emergencies. PloS one. 2011;6(3):e17680.
31. Hughes AL, Palen L. Twitter adoption and use in mass convergence and emergency events. International Journal of Emergency Management. 2009;6(3-4):248-60.
32. Li J, Rao HR. Twitter as a rapid response news service: An exploration in the context of the 2008 China earthquake. The Electronic Journal of Information Systems in Developing Countries. 2010;42.
33. Van Hentenryck P, editor Computational Disaster Management. IJCAI; 2013.
34. Wang D, Lin Y-R, Bagrow JP. Social networks in emergency response.  Encyclopedia of Social Network Analysis and Mining: Springer; 2014. p. 1904-14.
35. Watts D, Cebrian M, Elliot M. Dynamics of social media. Public Response to Alerts and Warnings Using Social Media: Report of a Workshop on Current Knowledge and Research Gaps. The National Academies Press, Washington, DC; 2013.
36. Caragea C, McNeese N, Jaiswal A, Traylor G, Kim H-W, Mitra P, et al., editors. Classifying text messages for the haiti earthquake. Proceedings of the 8th international conference on information systems for crisis response and management (ISCRAM2011); 2011: Citeseer.
37. Earle PS, Bowden DC, Guy M. Twitter earthquake detection: earthquake monitoring in a social world. Annals of Geophysics. 2012;54(6).
38. Freeman M. Fire, wind and water: Social networks in natural disasters. Journal of Cases on Information Technology (JCIT). 2011;13(2):69-79.
39. Guy M, Earle P, Ostrum C, Gruchalla K, Horvath S, editors. Integration and dissemination of citizen reported and seismically derived earthquake information via social network technologies. International Symposium on Intelligent Data Analysis; 2010: Springer.
40. Pickard G, Pan W, Rahwan I, Cebrian M, Crane R, Madan A, et al. Time-critical social mobilization. Science. 2011;334(6055):509-12.
41. Sakaki T, Okazaki M, Matsuo Y, editors. Earthquake shakes Twitter users: real-time event detection by social sensors. Proceedings of the 19th international conference on World wide web; 2010: ACM.

42. Skinner J. Natural disasters and Twitter: Thinking from both sides of the tweet. First Monday. 2013;18(9).
43. Ukkusuri S, Zhan X, Sadri A, Ye Q. Use of social media data to explore crisis informatics: Study of 2013 Oklahoma tornado. Transportation Research Record: Journal of the Transportation Research Board. 2014;(2459):110-8.
44. Abbasi A, Rashidi TH, Maghrebi M, Waller ST, editors. Utilising Location Based Social Media in Travel Survey Methods: bringing Twitter data into the play. Proceedings of the 8th ACM SIGSPATIAL International Workshop on Location-Based Social Networks; 2015: ACM.
45. Maghrebi M, Abbasi A, Rashidi TH, Waller ST, editors. Complementing Travel Diary Surveys with Twitter Data: Application of Text Mining Techniques on Activity Location, Type and Time. 2015 IEEE 18th International Conference on Intelligent Transportation Systems; 2015: IEEE.
46. Hasan S, Ukkusuri SV. Urban activity pattern classification using topic models from online geo-location data. Transportation Research Part C: Emerging Technologies. 2014;44:363-81.
47. Hasan S, Ukkusuri SV. Location contexts of user check-ins to model urban geo life-style patterns. PloS one. 2015;10(5):e0124819.
48. Zhao S, Zhang K, editors. Observing Individual Dynamic Choices of Activity Chains From Location-Based Crowdsourced Data. Transportation Research Board 95th Annual Meeting; 2016.
49. Cebelak MK. Location-based social networking data: doubly-constrained gravity model origin-destination estimation of the urban travel demand for Austin, TX. 2013.
50. Chen Y, Mahmassani HS, editors. Exploring Activity and Destination Choice Behavior in Two Metropolitan Areas Using Social Networking Data. Transportation Research Board 95th Annual Meeting; 2016.
51. Jin P, Cebelak M, Yang F, Zhang J, Walton C, Ran B. Location-Based Social Networking Data: Exploration into Use of Doubly Constrained Gravity Model for Origin-Destination Estimation. Transportation Research Record: Journal of the Transportation Research Board. 2014;(2430):72-82.
52. Lee JH, Gao S, Goulias KG, editors. Comparing the Origin-Destination Matrices from Travel Demand Model and Social Media Data. Transportation Research Board 95th Annual Meeting; 2016.
53. Yang F, Jin PJ, Wan X, Li R, Ran B, editors. Dynamic origin-destination travel demand estimation using location based social networking data. Transportation Research Board 93rd Annual Meeting; 2014.
54. Collins C, Hasan S, Ukkusuri SV. A novel transit rider satisfaction metric: Rider sentiments measured from online social media data. Journal of Public Transportation. 2013;16(2):2.
55. Carrasco J-A, Miller EJ. The social dimension in action: A multilevel, personal networks model of social activity frequency between individuals. Transportation Research Part A: Policy and Practice. 2009;43(1):90-104.
56. Sadri AM, Lee S, Ukkusuri SV. Modeling Social Network Influence on Joint Trip Frequency for Regular Activity Travel Decisions. Transportation Research Record: Journal of the Transportation Research Board. 2015;(2495):83-93.
57. Scott J. Social network analysis: Sage; 2012.
58. Alstott J, Bullmore E, Plenz D. powerlaw: a Python package for analysis of heavy-tailed distributions. PloS one. 2014;9(1):e85777.

59. Malevergne Y, Pisarenko V, Sornette D. Gibrat's law for cities: uniformly most powerful unbiased test of the Pareto against the lognormal. Swiss Finance Institute Research Paper. 2009;(09-40).
60. Malevergne* Y, Pisarenko V, Sornette D. Empirical distributions of stock returns: between the stretched exponential and the power law? Quantitative Finance. 2005;5(4):379-401.
61. Granovetter MS. The strength of weak ties. American journal of sociology. 1973:1360-80.
62. Miritello G, Moro E, Lara R. Dynamical strength of social ties in information spreading. Physical Review E. 2011;83(4):045102.

**TABLE 1:** Description of the Tweets and Network Elements

| | |
|---|---:|
| *Description of the Tweets* | |
| Number of Total Tweets | 56,159 |
| Number of Tweets without any User Mentions | 19,532 |
| Number of Tweets with at least one User Mentions | 36,627 |
| Number of Tweets only including Self Mentions | 20,645 |
| Number of Words | 3,589,732 |
| *Description of Network Elements* | |
| Number of Nodes (directed) | 34,363 |
| Number of Links (directed) | 39,709 |
| Network Density (directed) | 0.00003 |
| Number of Nodes (undirected) | 34,363 |
| Number of Links (undirected) | 38,442 |
| Network Density (undirected) | 0.00007 |
| Number of Nodes (lagest connected component) | 21,045 |
| Number of Links (lagest connected component) | 33,020 |
| Network Density (lagest connected component) | 0.00015 |
| Radius (lagest connected component) | 9 |
| Diameter (lagest connected component) | 17 |
| Number of Connected Components | 8,348 |
| Number of Isolates | 6,096 |
| Average Degree (directed) | 1.156 |
| Average Clustering Coefficient (undirected) | 0.149 |

Sadri, Hasan, Ukkusuri, Suarez Lopez 15**FIGURE 1: Description of the Tweet Database.** (a) Snapshot of 100 most frequent words in the dataset, (b) Tweets collected over time (Linked tweets versus Non-linked tweets)

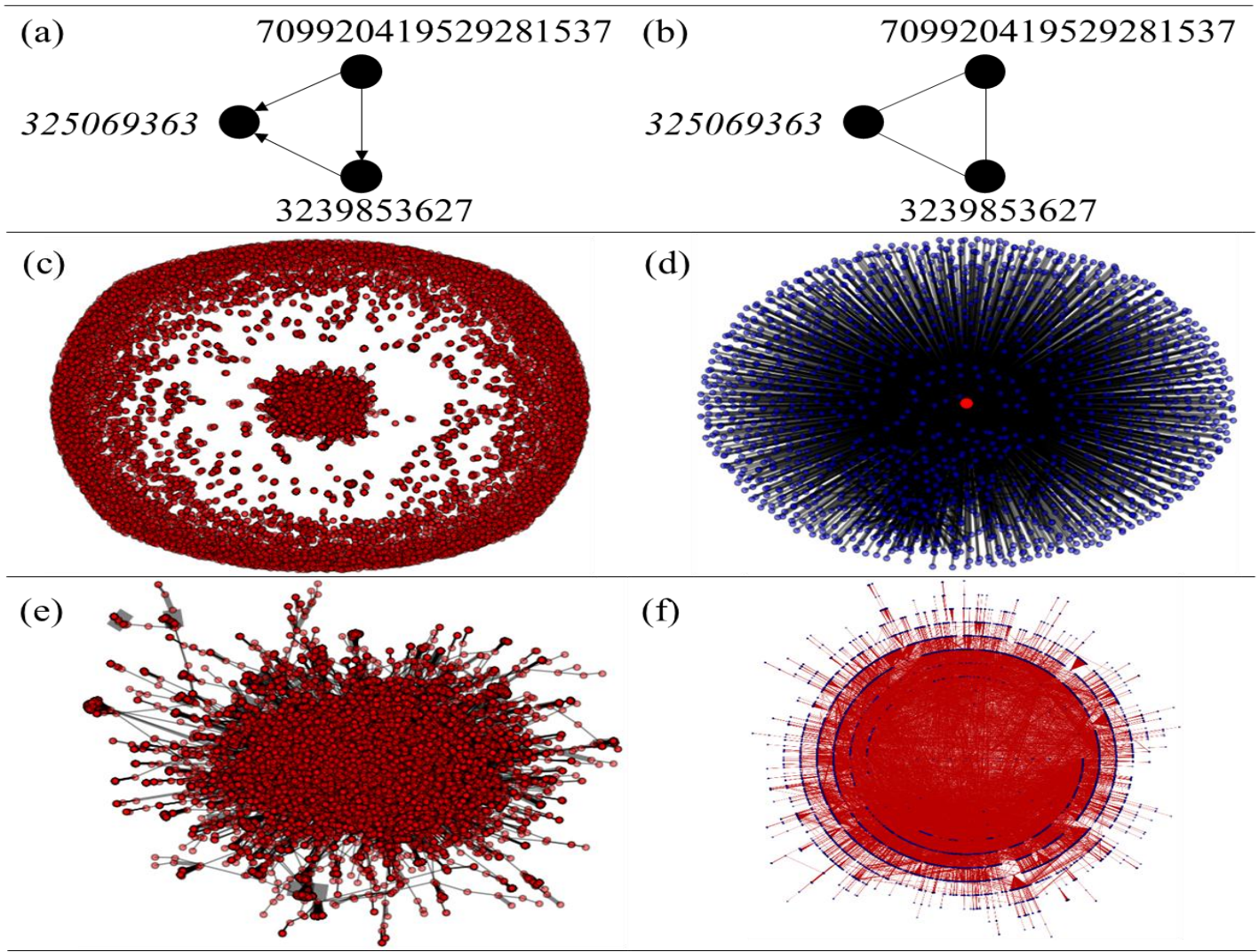

**FIGURE 2: Construction and Visualization of Social Interaction Network.** (a) Directed graph, (b) Undirected graph, (c) Undirected graph visualization, (d) Largest hub, (e) Weighted edges of the largest connected component, (f) Circular tree visualization of the largest connected component

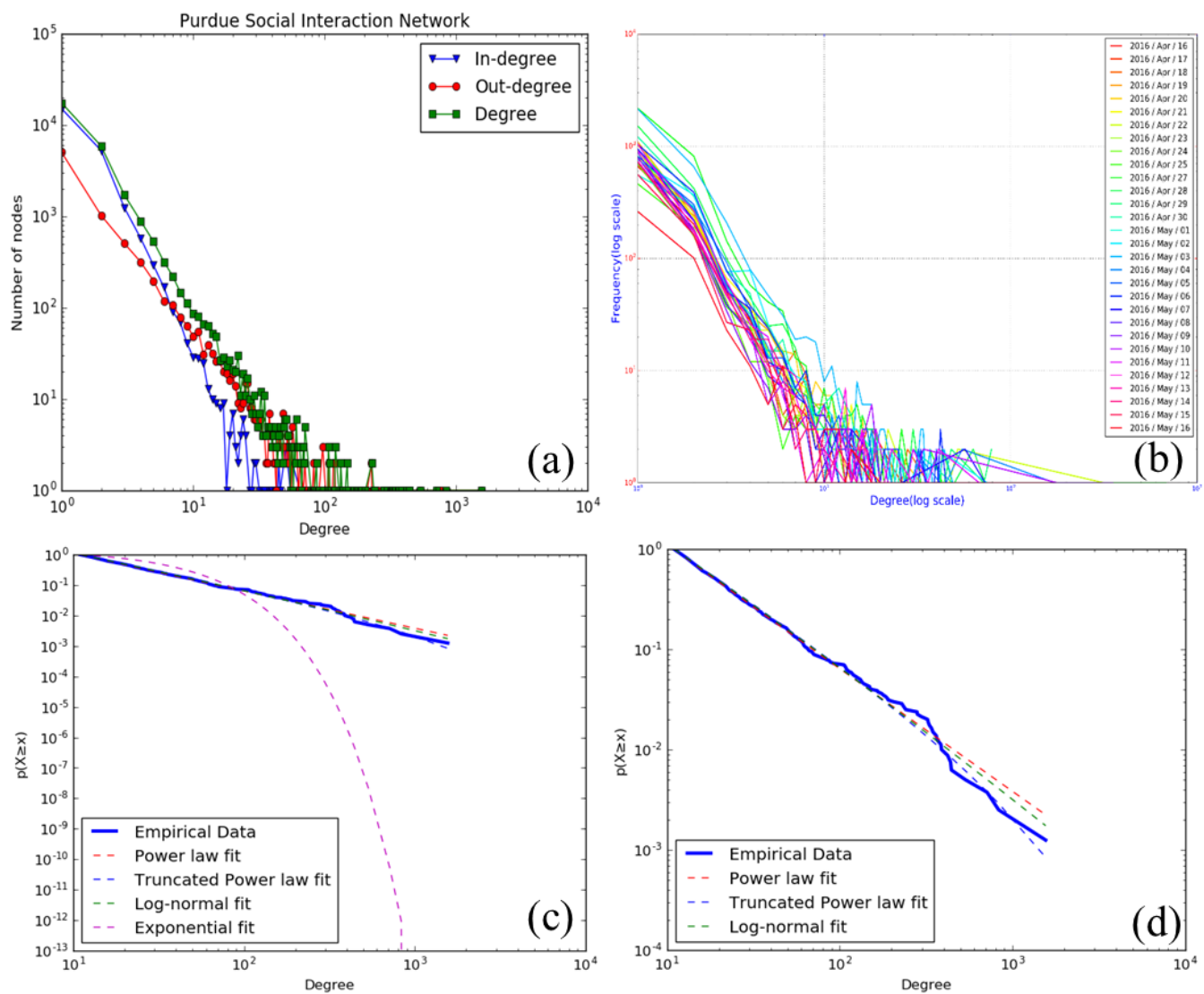

**FIGURE 3: Degree Distributions of Purdue Twitter Mention Network.** (a) In-degree, Out-degree and Degree Distributions (b) Degree distributions each day, (c) Comparison of data fitting with different distributions (d) Closer snapshot to power-law, truncated power-law and log-normal fitting comparisons

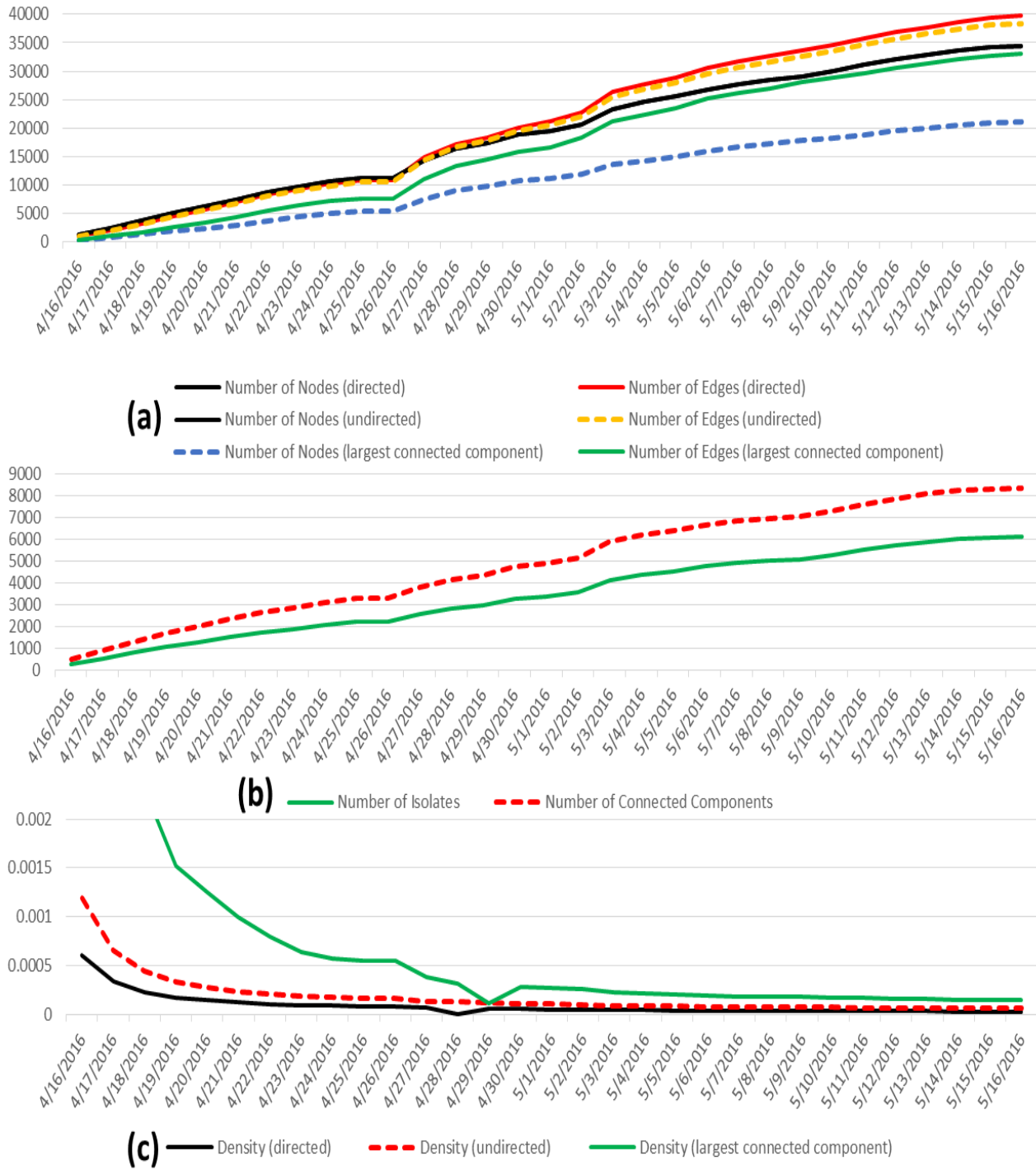

**FIGURE 4: Growth of Social Interaction Network over Time.** (a) Cumulative amount of network elements (nodes and edges), (b) Cumulative network isolates and connected generated each day, (c) Change in network densities each day due to additional elements.

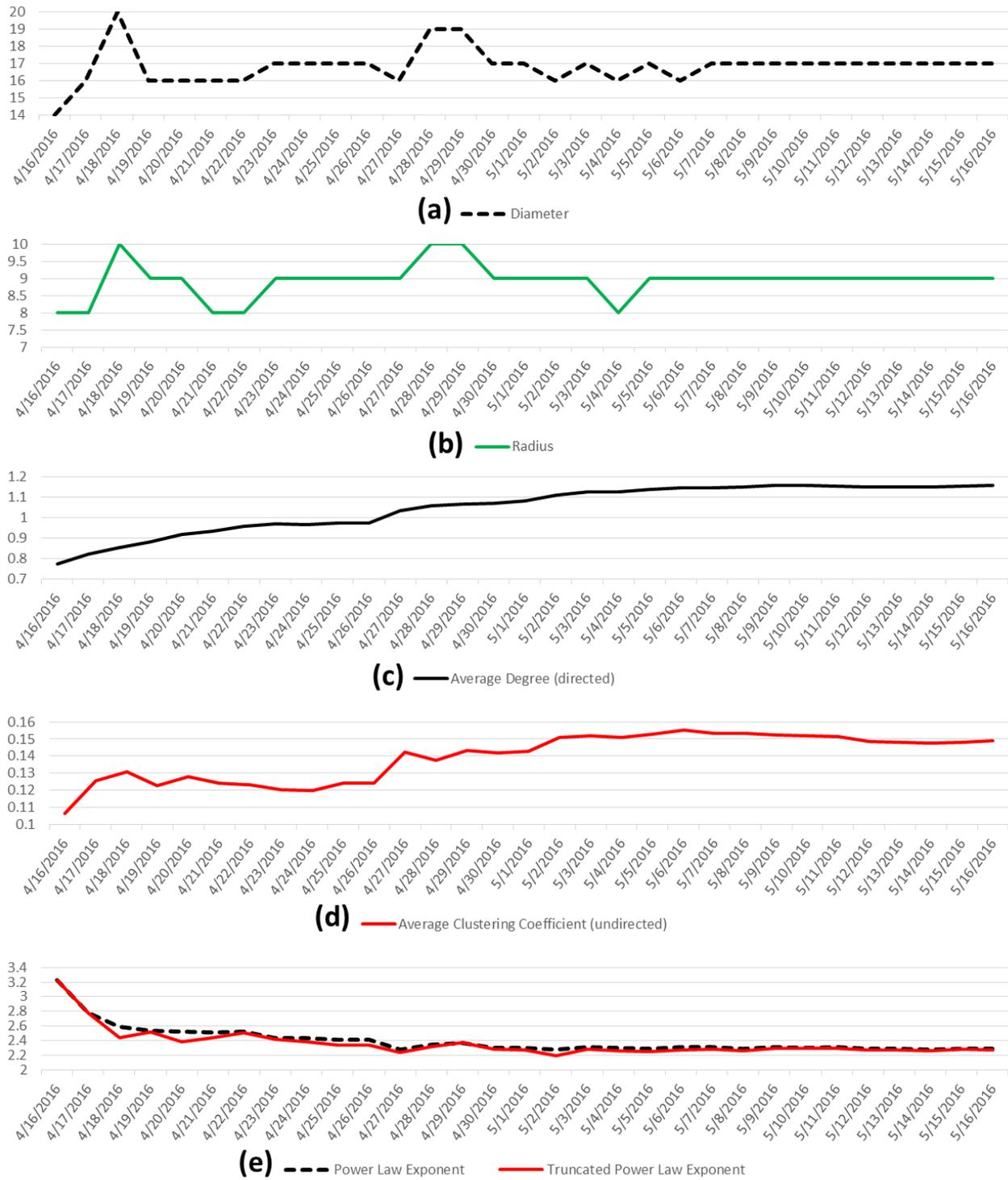

**FIGURE 5: Change in Social Interaction Network Properties over Time.** (a) Diameter, (b) Radius, (c) Average Degree, (d) Average Clustering Coefficient, (e) Power-law and Truncated Power-law Exponents.

Sadri, Hasan, Ukkusuri, Suarez Lopez 20

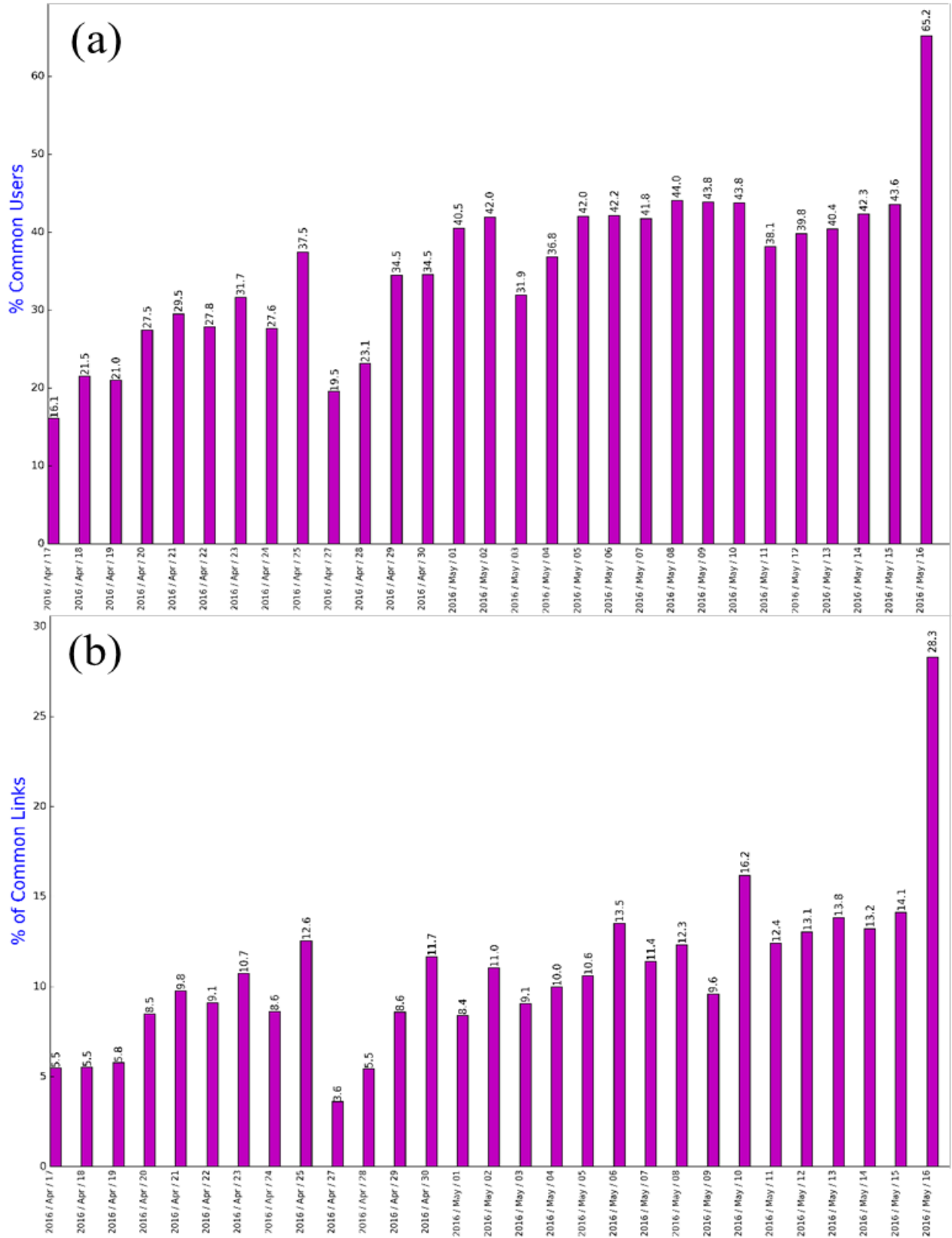

**FIGURE 6: Existence of Common Nodes and Links Each Day as compared to All Previous Days.** (a) Common users, (b) Common links (undirected)